\documentstyle[epsfig,graphicx,amssymb,natbib,psfrag]{jfm}

\title{Statistical mechanics of two-dimensional turbulence}

\author{Sunghwan Jung, P.J. Morrison$^1$, and Harry L. Swinney}
\affiliation{Center for Nonlinear Dynamics and Department of Physics,\\
$^1$Institute for Fusion Studies and Department of Physics, \\
The University of Texas at Austin, Austin, Texas 78712 USA.}
\pubyear{2006} \volume{600} \pagerange{1-20}
\date{\today}

\begin{document}
\maketitle
\begin{abstract}

The statistical mechanical description of two-dimensional inviscid
fluid turbulence is reconsidered.  Using this description, we make
predictions about turbulent flow in a rapidly rotating laboratory
annulus.  Measurements on the continuously forced, weakly dissipative
flow reveal coherent vortices in a mean zonal flow. Statistical
mechanics has two crucial requirements for equilibrium: statistical
independence of macro-cells (subsystems) and additivity of invariants
of macro-cells. We investigate these requirements in the context of
the annulus experiment.  The energy invariant, an extensive quantity,
should thus be additive; i.e., the interaction energy between a
macro-cell and the rest of the system (reservoir) should be small, and
this is verified experimentally.  Similarly, we use additivity to
select the appropriate Casimir invariants from the infinite set
available in vortex dynamics, and we do this in such a way that the
exchange of micro-cells within a macro-cell does not alter an
invariant of a macro-cell.  A novel feature of the present study is
our choice of macro-cells, which are continuous phase space curves
based on mean values of the streamfunction.  Quantities such as energy
and enstrophy can be defined on each curve, and these lead to a local
canonical distribution that is also defined on each curve.  The
distribution obtained describes the anisotropic and inhomogeneous
properties of a flow.  Our approach leads to the prediction that on a
mean streamfunction curve there should be a linear relation between
the ensemble-averaged potential vorticity and the time-averaged
streamfunction, and our laboratory data are in good accord with this
prediction.  Further, the approach predicts that although the
probability distribution function for potential vorticity in the
entire system is non-Gaussian, the distribution function of
micro-cells should be Gaussian on the macro-cells, i.e., for curves
defined by mean values of the streamfunction.  This prediction is also
supported by the data.  While the statistical mechanics approach used
was motivated by and applied to experiments on turbulence in a
rotating annulus, the approach is quite general and is applicable to a
large class of Hamiltonian systems, including drift-wave plasma
models, Vlasov-Poisson dynamics, and kinetic theories of stellar
dynamics.
\end{abstract}

\section{ Introduction }
\label{sec:Intro}

\subsection{Overview}

Statistical mechanics provides a way to calculate the macroscopic properties of
matter from the behavior of microscopic constituents.  Instead of considering
all motions of the individual constituents, one describes observable quantities
averaged over constituent Hamiltonian trajectories, and averages are evaluated
using the probability distribution of possible microstates. Likewise, fluid
systems with a local balance between dissipation and forcing have been
described by statistical mechanics with the inclusion of constraints based on
invariants of the dynamics.  In general, such statistical theories for fluids
are based on the idea that the macroscopic behavior of the fluid turbulence can
be described without knowing detailed information about small scale vortices.

The justification of statistical mechanics based on ideal two-dimensional fluid
equations is open to question, given the existence of forcing, dissipation,
three-dimensional effects, temperature gradients, etc., that certainly occur in
real fluid flows. Moreover, one must square the idea of cascading with the
approach to statistical equilibrium.  Ultimately, such a justification is very
difficult and would rely on delicate mathematical limits.  However, its success
amounts to the idea that the fluid system can in some sense be described by
weakly interacting subsystems, where the behavior of a single subsystem can be
described by weak coupling to a heat bath that embodies all of the other
subsystems and all of the omitted effects.  In the end `the proof of the
pudding is in the eating', and our justification is based on experimental
observations.

Intimately related to the existence of subsystems is the question of which
invariants to incorporate into a statistical mechanics treatment of fluids.  One
aim of the present paper is to investigate this question.  We investigate this
question both conceptually  and experimentally and come to the conclusion that
quadratic invariants (energy and enstrophy) are most important.  Our conclusion
follows from the observation that these invariants possess the property of
additivity.

The microscopic dynamics of conventional statistical mechanics is
finite dimensional, but to describe macrosopic phenomena one takes the
thermodynamic limit in which the number of degrees of freedom tends to
infinity.  However, the dynamics of a two-dimensional fluid is already
infinite dimensional and possesses an infinite number of invariants;
so, in order to make progress with a statistical mechanics approach
one must extract a finite-dimensional model, and such a model cannot
conserve all of the invariants of the original fluid system.  In
calculations one may also take limits of this finite-dimensional
model, but the results of these limits may depend upon which of the
invariants are maintained. Additivity of macroscopic invariants and
statistical independence of subsystems are crucial properties in
conventional statistical mechanics [see e.g.\ \cite*{landau_ST}].
Because not all invariants of a system are additive, this property can
be used to select invariants for statistical mechanics from the
infinite number possessed by two-dimensional fluid systems.

Related to the choice of additive invariants is the choice of subsystems.  This
choice requires the identification of two scales, a macroscopic scale and a
microscopic scale, which we call $\Delta$ and $\delta$, respectively, and phase
space cells of these characteristic sizes are considered.  In classical
statistical mechanics, the micro-cells usually refer to individual particles,
while the macro-cells, the subsystems, are selected to be large enough to
contain many particles yet small enough to have uniform invariants.  We address
in detail the choice of these cells for the fluid in \S\ref{sec:MF}, but it is
clear that a macro-cell should contain many micro-cells, yet be small enough so
that the vorticity and streamfunction are constant.  This condition is
sufficient for statistical independence, but the converse is not always true.
In any event we seek to define macro-cells that are nearly statistically
independent and consider only invariants that are additive over these cells.

A second aim of the present work is to propose the idea that temporal mean
values of the streamfunction provide a natural coordinate system for describing
inhomogeneous turbulence, a coordinate system that can be used to define
statistically independent subsystems.  We suggest this idea because contours of
the streamfunction for two-dimensional inviscid fluid flow tend to be smooth
and because there tends to be a strong statistical dependence of vorticity or
potential vorticity along those contours.  Streamfunction contours are much
smoother than vorticity contours because of the smoothing property of the
inverse Laplacian.  Therefore, there is a natural separation of length scales:
the large scale associated with variation of the streamfunction contours and
the fine scale that is needed to resolve the vorticity or potential vorticity.
We take these to be our scales $\Delta$ and $\delta$, respectively.  We test
this idea experimentally by measuring the independence of subsystems so
defined. We then construct a theory based on this definition of subsystem
together with the additivity of quadratic invariants, and compare its
predictions with the measured vorticity probability density function.

\subsection{Background}

In a remarkable series of papers \cite*{burgers29a,burgers29b,
burgers29c,burgers33a,burgers33b,burgers33c,burgers33d} [reprinted in
\cite*{burgers95}] appears to be the first researcher to apply
statistical mechanics ideas to the description of fluid turbulence.
Many basic ideas used by later researchers were introduced first by
Burgers in these rarely cited papers.  Burgers introduced both lattice
and Fourier models and showed that such models satisfy Liouville's
theorem when viscosity is neglected.  He used a counting argument to
derive an entropy expression and obtained a corresponding entropy
maximization principle.  He proposed a microscopic scale for
describing turbulent motion during short intervals of time and defined
macroscopic quantities by counting possible streamfunction
realizations for sequences of time intervals.  His analysis is based
on the Reynolds stress equation, and he obtained a probability
distribution that can be used to calculate the mean value of the
Reynolds stress.

Motivated by the work of Burgers, \cite*{onsager49} took up the
subject and considered a representation of the vorticity field in
terms of a set of point vortices, zero-area vortices, of equal
strength.  Because this results in a finite-dimensional particle-like
Hamiltonian system, Onsager could proceed to apply techniques of
classical statistical mechanics.  He gave arguments for the existence
of negative temperatures and the occurrence of coherent structures in
a confined region, which are often observed in nature.  Related ideas
have been further pursued by many researchers [e.g.\
\cite*{joyce73,matthaeus91,eyink93,ST04}] [see \cite*{eyink05} for a
recent review].  For example, \cite*{joyce73} studied the statistical
mechanics of point vortices within a mean field approximation, and
argued that in the negative temperature regime, large like-signed
vortices are the most probable state.

T.D.\ \cite*{lee52} projected three-dimensional fluid equations
(including MHD) onto a Fourier basis and truncated to obtain a
finite-dimensional system.  Evidently unaware of the early work of
\cite*{burgers33d}, he again demonstrated that his truncated system
satisfies a version of Liouville's theorem and was thus amenable to
techniques of statistical mechanics.  Later, Kraichnan considered
two-dimensional fluids [\cite*{kraichnan67,kraichnan75,kraichnan80}]
and noted that out of the infinite number of invariants, two quadratic
invariants, the so-called rugged invariants, remained invariants after
truncation. Kraichnan and Montgomery argued 
that these rugged invariants are the important ones, and obtained an
equilibrium state, which is related to that obtained by minimum
enstrophy arguments put forth by selective decay hypotheses
[\cite*{leith84,maxworthy84,ST22}].  Also, using Kolmogorov-like
dimensional arguments and the rugged invariants, Kraichnan argued for
the existence of direct and inverse cascades for two-dimensional
turbulence [\cite*{kraichnan67}].

The two-dimensional Euler equation, like the Vlasov and other
transport equations, can be viewed as mean field theory.  Such
equations are known to generate fine structure in the course of
evolution.  This led \cite*{lynden67} to consider a coarse graining
procedure coupled with the idea of preserving all of the infinity of
invariants such theories possess.  He applied his ideas in the context
of stellar dynamics, but the ideas are akin to those used in
treatments of the classical electron gas by generalizations of
Debye-H\"uckle theory [e.g.\ \cite*{vankampen67}].  Later, such ideas
were used in the fluid context by \cite*{robert91b}, \cite*{robert91},
\cite*{robert92}, \cite*{miller90}, and \cite*{miller92}, and again in
the stellar dynamics context by \cite*{ST132}.  Our development to a
large degree parallels that of these authors.  In these works a
microscopic probability distribution represents a local description of
the small-scale fluctuations of microscopic vortices.  The
streamfunction is assumed to be uniform on the microscopic scale, and
an equilibrium state is obtained by maximizing the Boltzmann entropy
of microstates, an entropy that is obtained by a counting argument
first given by Lynden-Bell.  This produces a most probable state.

More recently, the necessity of incorporating the infinite number of
invariants in statistical mechanics theories has been brought into
question, and theories based on finite-dimensional models with fewer
constraints have been developed. \cite*{ST112} have argued that
including an infinite number of invariants provides no additional
statistical information, and \cite*{ST74} has argued that previous
theories have not properly handled the neglected small scale
phenomena, and he has proposed a theory that uses inequality
constraints associated with only the convex invariants.  Our approach
is perhaps most closely aligned to these works, but is distinguished
by the fact that the invariants chosen are explicitly based on the
additivity argument, the choice of subsystems, and experimental observation.

Natural phenomena in atmospheres and oceans have served as a
motivation for the application of statistical mechanics to
two-dimensional fluid flow [e.g.\ \cite*{salmon76}].  Examples include
zonal flows in planets, such as the jet stream and the polar night
jet, and organized coherent vortices, such as the Great Red Spot of
Jupiter [\cite*{maxworthy84,som91,sommeria88,marcus93,ST22}].
Attempts have been made to explain such naturally occuring phenomena
in terms of the coherent structures found to emerge in
quasi-geostrophic and two-dimensional turbulence after long time
evolution.  With external small-scale forcing a few long-lived and
large structures resulting from nonlinear merging processes are seen
to be stable self-organized states that persist in a strongly
turbulent environment [\cite*{mcwilliams84,ST61}]. These structures
have been studied over many years, often because of their relevance to
large-scale geophysical and astrophysical flows [\cite*{marcus93}].
In statistical mechanics, such steady states with large structures are
envisioned to be the most probable state arising from some
extremization principle.  Various extremization principles [e.g.\
\cite*{leith84}] have been proposed with selected global invariants of
the system used as constraints. Observations of turbulent flow with
large coherent structures in a rotating annulus
[\cite*{sommeria88,baroud02,baroud03,aubert02,jung04}] have led us to
reconsider statistical mechanics in the context of rapidly rotating
systems.

\subsection{Notation \& Organization}

By necessity this paper contains much notation. To aid the reader we give a
brief summary here.  As noted above, statistical mechanics deals with two
scales: the microscopic scale $\delta$, characteristic of microscopic
$m$-cells, and the macroscopic scale $\Delta$, characteristic of macroscopic
$M$-cells. Several averages are considered.  The symbol $\langle \: \cdot \:
\rangle_{S}$ denotes an average with probability density $P_{S}$, where choices
for the subscript $S$ will be used to delineate between different cases.  The
appropriate volume measure will be clear from context but is also revealed by
the argument of $P_{S}$. Averages with uniform density are denoted by $\prec
\cdot \! \succ_{S}$, where the subscript denotes the integration variable.  An
exception is the time average, which we denote by an overbar. Thus, the time
average of a function is denoted by $\bar f$, and $\bar f = \int_{0}^{T} f dt/T
= \prec \! f \! \succ_{t}$. The limits of integration for this kind of average
will either be stated or will be clear from context.  We denote the potential
vorticity field by $q(x,y,t)$, by which we always mean a function.  For the
potential vorticity distribution on a $M$-cell (subsystem) we use $\zeta$, an
independent variable.  Another source of possible confusion is that the symbol
$\beta$ is used for the energy Lagrange multiplier, as is conventional in
statistical mechanics, while the beta-effect of geophysical fluid dynamics is
embodied here in the symbol $h$.

The paper is organized as follows.  The experiment is described in
\S\ref{sec:experiment} and equations that govern the dominant physics are
reviewed in \S\ref{sec:dynamics}.  In \S\ref{sec:SM} we describe some basic
ideas about statistical mechanics, as needed for the application to the fluid
system of interest.  In \S\ref{sec:MF} we describe statistical mechanics in the
mean field approximation and compare predictions with experiments.  Here we
show that predictions of the theory are in accordance with experiments.
Finally, in \S\ref{sec:discuss} we conclude.

\section{Experiment}

\label{sec:experiment}
\begin{figure} 
\begin{center} \psfrag{qs}{$\omega$ (s$^{-1}$)} 
\psfrag{(a)}{\bf (a)} \psfrag{(b)}{\bf (b)} \psfrag{0}{0} \psfrag{ro}{$r_o$}
\psfrag{ri}{$r_i$} \psfrag{VthetaVtheta}{$\prec \! \overline{ U_{\theta}} \!
\succ_{\theta}$ (cm/s)} \psfrag{radius}{radius (cm)} \psfrag{omega}{$\prec \!
\overline{ \omega} \! \succ_{\theta}$ (s$^{-1}$)} \psfrag{(c)}{\bf (c)}
\psfrag{(d)}{\bf (d)} \psfrag{H}{\hspace{-.8mm}$L_h$}
\psfrag{psipsi}{\vspace{9mm}$\prec \! \bar \psi  \! \succ_{\theta}$ (cm$^2$/s)}
\psfrag{X2}{$\prec \! \overline{ \omega} \! \succ_{\theta}$} \psfrag{X1}{$\prec
\! \bar \psi  \! \succ_{\theta}$}
\includegraphics[width=.5\linewidth]{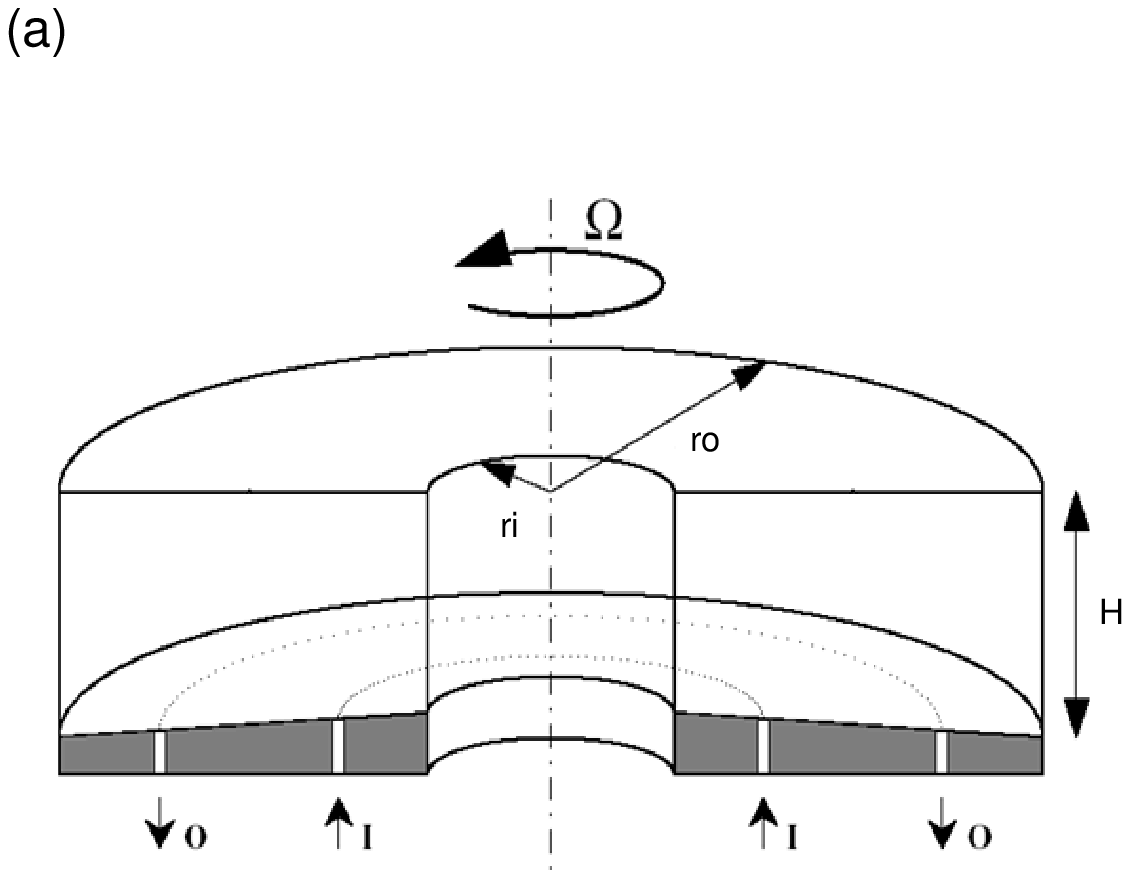}
\hspace{-0.2in}
\includegraphics[width=.5\linewidth]{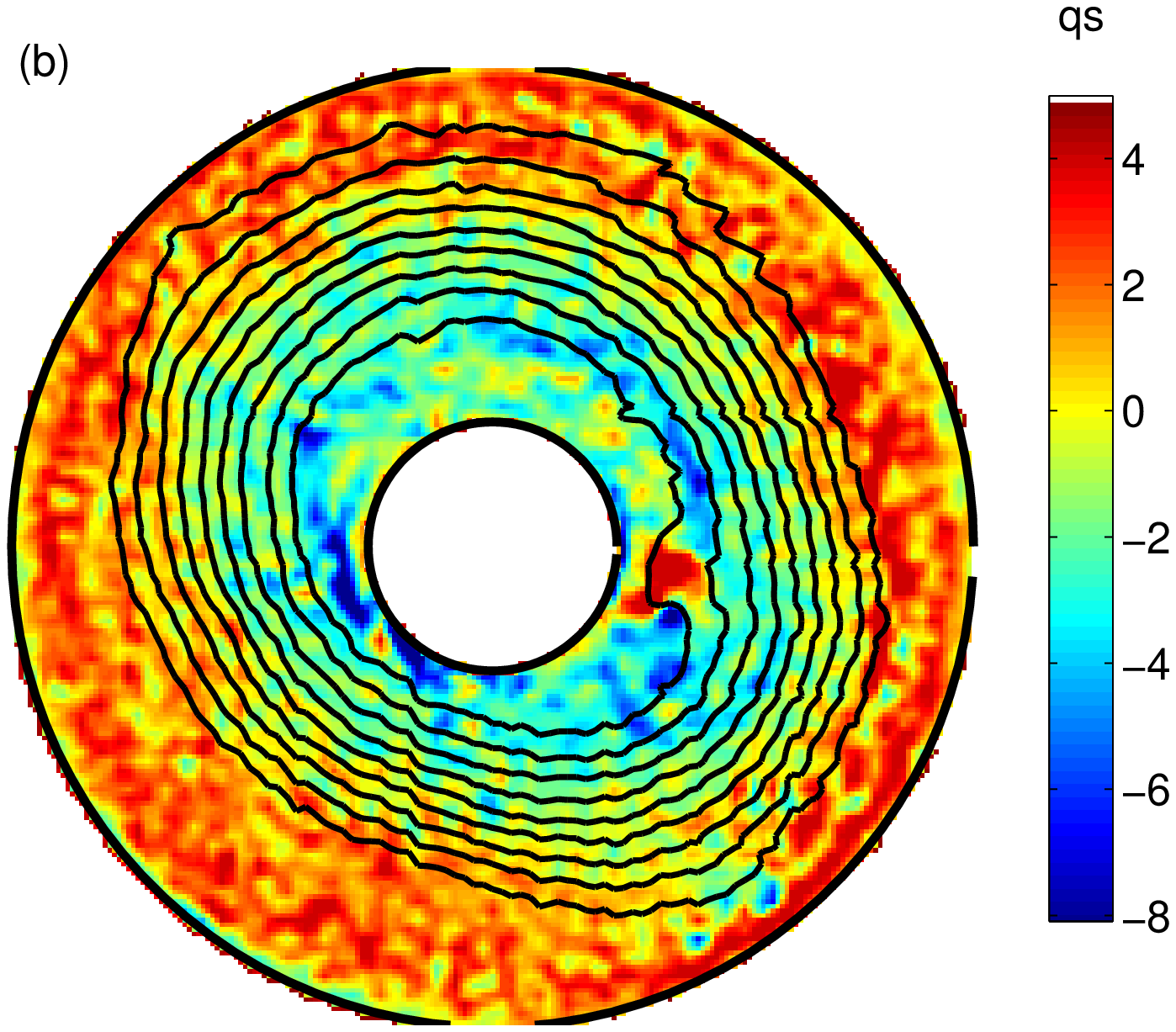} \\
\includegraphics[width=\linewidth] {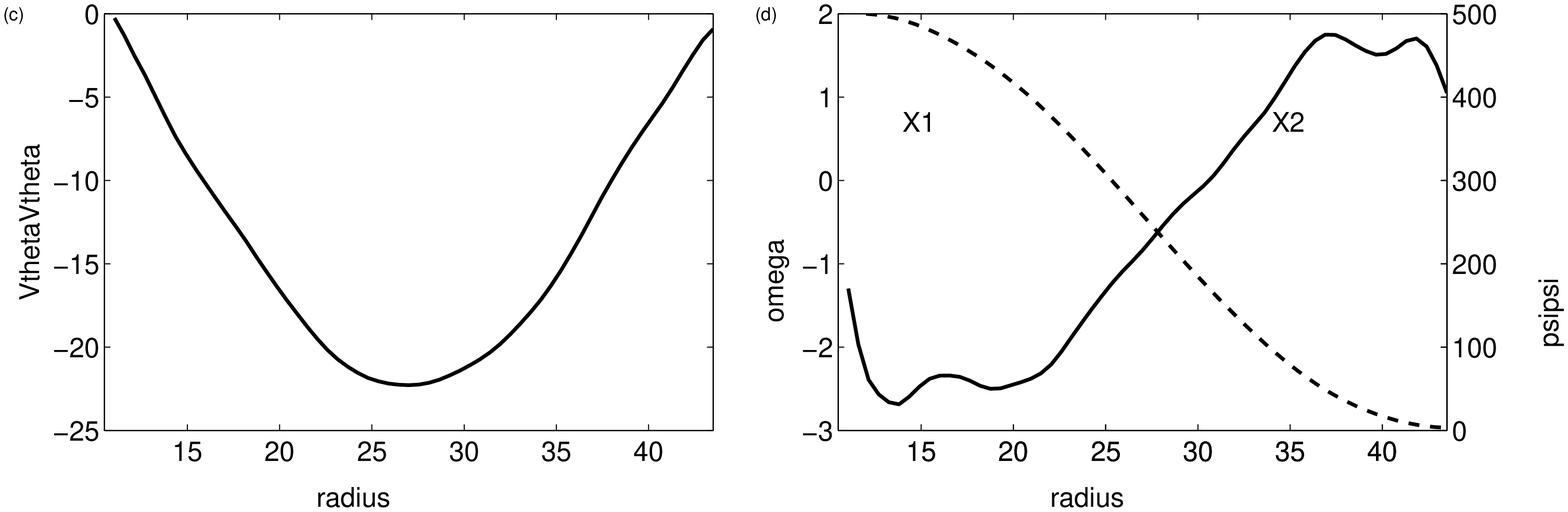}
\end{center}
\caption{(a) Schematic diagram of the experimental apparatus.  The tank rotates
at 1.75 Hz. Flow is produced by pumping water through a ring of inlets (I) and
outlets (O) in the bottom of the tank. The Coriolis force acts on the radially
pumped fluid to produce a counter-rotating jet.  (b) The vorticity field and
contours of streamfunction at mid-height of the tank, determined from Particle
Image Velocimetry measurements. The streamfunction contours are equally spaced
in streamfunction value.  (c) The azimuthal velocity averaged over both time
and azimuthal angle, as a function of radial position.  (d) The vorticity
(solid line) and streamfunction (dashed line) averaged over time and azimuthal
angle, as a function of radial position. } \label{fig:apparatus}
\end{figure}

The experiments are conducted in a rotating annulus (Fig.~\ref{fig:apparatus}).
The annulus has an inner radius $r_i=10.8 \mathrm{~cm}$, outer radius $r_o
=43.2 \mathrm{~cm}$, a sloping bottom, and a flat transparent lid.  The bottom
depth varies from $17.1 \mathrm{~cm}$ at the inner radius to $20.3
\mathrm{~cm}$ at the outer radius, giving a bottom slope of $\eta = -0.1$.  For
the data analyzed in this paper, the rotation frequency of the annulus is
$\Omega/2\pi= 1.75$ Hz.  An azimuthal jet is generated in the annulus by
pumping water in a closed circuit through two concentric rings of holes at the
bottom. Fluid is pumped into the annulus through an inner ring at $r=18.9
\mathrm{~cm}$ and extracted through an outer ring at $r=35.1 \mathrm{~cm}$;
both rings have 120 circular holes.  Each hole has a diameter of $2.5
\mathrm{~mm}$, and the total pumping rate is 150 $\mathrm{cm}^3/\mathrm{s}$.
The action of the Coriolis force on the outward flux generates a
counter-rotating azimuthal jet. A counter-rotating flow is generally more
unstable than a co-rotating flow [\cite*{som91}].

The water is seeded with neutrally buoyant particles (polystyrene spheres,
diameter $150-200 \mathrm{~\mu m}$).  Light emitting diodes produce a $3
\mathrm{~cm}$ thick horizontal sheet of light that illuminates the annulus at
mid-depth.  The particles suspended in the water are imaged with a camera
located $2 \mathrm{~m}$ above the annulus, and the camera rotates with the
tank. Particle Image Velocimetry (PIV) is used to obtain the full
two-dimensional velocity field [\cite*{baroud03}].

The flow can be characterized by the Reynolds, Rossby, and Ekman
numbers.  The maximum velocity $U_{\mathrm{max}} \approx 22$ cm/s, the
length $L=16.2$ cm (taken to be the distance between the two forcing
rings) and the kinematic viscosity $\nu = 0.01$ cm$^2$/s yield a
Reynolds number $UL/\nu=3.5 \times 10^4$, indicating that the flow is
turbulent. The Rossby number (ratio of inertial to Coriolis force) is
${\omega}_{rms}/2 \Omega = 0.11$ (where ${\omega}_{rms}$ is the rms
vorticity), which indicates that the Coriolis force is dominant, as is
the case for planetary flows on large length scales.  Finally, the
small Ekman number, $\nu/2 L^2 \Omega = 3\times10^{-4} $, indicates
that dissipation in the bulk is small.  The Ekman time, $\tau_E =
L_h/2(\nu \Omega)^{1/2}$ (where $L_h$ is the mean fluid height) for
dissipation in the boundary layers is $30 \mathrm{~sec}$, a time much
longer than the typical vortex turnover time, $2 \mathrm{~sec}$. The
dimensionless numbers indicate that the flow is quasi-geostrophic;
previous studies of turbulence in the annulus have indeed confirmed
the strong two-dimensionality of the flow [\cite*{baroud03}].

\section{Dynamics}
\label{sec:dynamics}

The barotropic assumption is widely used to describe the flow inside
the tank. The equation of motion for a barotropic fluid with
topography is given by
\begin{equation}
\frac{\partial q}{\partial t} + {\mathbf u} \cdot \nabla q = D + F \, ,
\label{eqmn}
\end{equation}
where $q = (-\nabla^2 \psi + 2 \Omega)/L_h$ is the potential vorticity, $L_h$
is the tank depth, $\psi$ is the streamfunction, ${\mathbf u} = (\partial
\psi/\partial y,-\partial \psi/\partial x)$,  $D$ denotes dissipation, such
as that due to molecular viscosity, $\nu \nabla^2 \omega$, or Ekman drag,
$-\omega/\tau_{E}$, and $F$ denotes a vorticity source due to the pumping.
Often the potential vorticity is approximated by
\begin{equation}
q = -\nabla^2 \psi + h\,,
\end{equation}
where $h$ accounts for the beta-effect and is here a linear function of radius,
$h= 2 \Omega \eta r/L_h$ where $\eta$ is the bottom slope.  Over the years
strong evidence has accumulted that (\ref{eqmn}) describes the dominant
features of the experiment [\cite*{som91,DCN92,meyers93,solomon93}].

For inviscid flow with zero Rossby number, there is no vertical variation in
the velocity [\cite{rossby39}], and there is evidence that to leading order the
drag and forcing terms cancel.  We are primarily interested in the statistics
of motions that occur on the vortex turnover time, and these are governed by
the inviscid equation,
\begin{equation}
\frac{\partial q}{\partial t} + {\mathbf u} \cdot \nabla q = 0,
\label{eq:euler}
\end{equation}
which is a Hamiltonian theory.

A manifestation of the Hamiltonian nature of two-dimensional Euler-like flows
such as (\ref{eq:euler}) is the finite-dimensional Hamiltonian description of
point vortices provided by \cite{kirchoff83}, which played an essential
motivating role in Onsager's theory [e.g.\ \cite{eyink05}].  For a distributed
vorticity variable such as $q$ the  Hamiltonian form is infinite-dimensional
and is given in terms of a noncanonical Poisson bracket as follows:
\begin{equation}
     \frac{\partial q}{\partial t} = \{q,H\}= [\psi,q]\,,
\end{equation}
where the Hamiltonian $H[q]=\int \psi (q-h) dxdy/2$, and the noncanonical
Poisson bracket is given by
\begin{equation}
\{F,G\}=\int q \left[\frac{\delta F}{\delta q}, \frac{\delta G}{\delta
q}\right]\, dxdy\,,
\label{ncbkt}
\end{equation}
with $F$ and $G$ being functionals, $\delta F/\delta q$ the functional
derivative, and $[f,g]=f_{x}g_{y}-f_{y}g_{x}$.  Observe that ${\mathbf u} \cdot
\nabla q= - [\psi,q]$.  This Hamiltonian formulation of the two-dimensional
Euler equation appeared in \cite{morr81,morr82}, based on the identical
structure for the Vlasov-Poisson system [\cite{morr80}], and in \cite{olver82}.
A review of this and other forumlations can be found in \cite{morr98}.  The
infinite family of Casimir invariants $C[q]=\int {\cal C}(q) dxdy$, where
${\cal C}$ is arbitrary, satisfies $\{F,C\}=0$ for all functionals $F$, and is
thus conserved by (\ref{eq:euler}).  The presence of these invariants is one
way that the statistical mechanics of fluids differs from that of particle
systems.

\section{Statistical Mechanics \& Fluid Mechanics}
\label{sec:SM}

As noted in \S\ref{sec:Intro}, many attempts have been made to apply statistical
mechanics to fluids and other infinite-dimensional systems.  In this section we
introduce our notation and discuss some basic ideas.

\subsection{State variables}
\label{ssec:state}

In classical statistical mechanics the microscopic dynamics is governed by
Hamilton's equations and the phase space is the $2N$-dimensional manifold with
canonical coordinates $(Q_{\alpha},P_{\alpha})$, $\alpha=1,2,\dots, N$, where
$(Q_1,\dots Q_N)$ is the configuration coordinate and $(P_1,\dots,P_N)$ is the
corresponding canonical momentum.  Typically $N$, the number of degrees of
freedom, is a very large number $\sim 10^{23}$.  We call this $2N$-dimensional
phase space $\Gamma$, a standard notation introduced by P.\ and T.\ Ehrenfest
[\cite*{Ehrenfest}].  Our fluid is assumed to be governed by (\ref{eq:euler}),
an infinite-dimensional Hamiltonian theory, and thus the instantaneous state of
our system is determined by the vorticity-like variable $q(x,y)$, which we
suppose is contained in some space of functions $\cal G$.  The index $\alpha$
for coordinates of $\Gamma$ is analogous to the Eulerian position $(x,y)$, a
point in the physical domain occupied by the fluid, which is viewed as an index
for $\cal G$.

In conventional statistical mechanics, the microscopic dynamics is
finite dimensional, and one attempts to explain phenomena on the
macroscopic level by considering the thermodynamic limit in which
$N\rightarrow\infty$.  However, for a fluid, the dynamics is already
infinite dimensional, and thus as noted in \S\ref{sec:Intro}, to apply
statistical mechanics researchers have introduced various
finite-dimensional discretizations.  Onsager's description of the
continuum vortex dynamics in terms of a collection of point vortices
amounts to the specification of the coordinates of the manifold
analogous to $\Gamma$ as the spatial positions of the point vortices,
$(x_1,\dots x_N,y_1,\dots,y_N)$.  Alternatively, Lee's representation
of a three-dimensional fluid in terms of a truncated Fourier series
has the Fourier amplitudes being coordinates of a space analogous to
$\Gamma$.  This procedure was carried over to two dimensions by
Kraichnan and Montgomery [\cite*{kraichnan80}].  For our potential
vorticity variable the Fourier amplitudes are given by $q_{\mathbf k}
= \int \exp{i(k_x x + k_y y)}\, q(x,y) dx dy$, where ${\mathbf
k}=(k_x,k_y)$.  Another alternative is to replace the continuum
vorticity by a lattice model [e.g.\ \cite*{burgers29a,robert91b,
robert91, robert92,miller90,miller92,ST112,ST74}], i.e., an expansion
in terms of tent functions or finite elements of scale size $\delta$.
In the present context the vorticity is replaced by its values on the
lattice, $q_i = \int K_i(x,y; x_i, y_i) q(x,y) dxdy$, where the kernel
$K_i$ is typically chosen to represent a square lattice with a finite
number $N$ of sites located at $(x_i,y_i)$.  In general $N= N_x N_y$,
where $N_x$ and $N_y$ are the number of lattice points in the $x$ and
$y$ directions, respectively. We will refer to this discretization as
a division into $m$-cells.

Given a finite-dimensional system one can make various assumptions, e.g., the
probabilistic assumptions of `molecular chaos', but this requires a notion of
phase space volume conservation.

\subsection{Phase Space Volume \& Liouville's Theorem}

In classical statistical mechanics one calculates averages over the manifold
$\Gamma$, and the natural volume element is given by $\Pi_{\ \alpha =1}^{N}
dQ_\alpha dP_\alpha$.  However, for $\cal G$ the situation is not so
straightforward, and so we explore candidates for the analogous volume element.

\subsubsection{Volume element}

The calculation of averages in a statistical theory requires a phase space
measure, ${\cal D}q$, which is a sort of volume element for $\cal G$.  The
volume element can be interpreted as a probability measure defined on functions
that take values between $q$ and $q + dq$.  Averages calculated using the
probability measure are functional integrals akin to those used in Feynman's
path integral formulation of quantum mechanics and in field theory [e.g.\
\cite*{schulman,sundermeyer}].  The various discretizations introduced above
have been employed to give meaning to functional integrals, but the Fourier and
lattice models are most common.

For the Fourier discretization, Kraichnan and Montgomery used the volume
element ${\cal D}q = \prod_{\mathbf{k}} d q_{\mathbf{k}}$, where the product is
truncated at some maximum wave number.  Alternatively, the volume element for
lattice models is written as ${\cal D}q = \prod_{i}^N dq_{i}$, where $dq_{i}$
is a volume element associated with the potential vorticity varying from $q$ to
$q+dq$ in a lattice partition ($x_i,y_i$), and $N= N_x N_y$ is, as above, the
number of lattice sites, which  have a scale $\delta$.  Here, a total volume
element ${\cal D}q$ is a product of volume elements of each lattice site
$dq_{i}$.  In the case of a finite small lattice, $dq_{i}$ becomes a 
one-dimensional volume, i.e., $dq_{i}= q(x_i,y_i) + dq(x_i,y_i) - q(x_i,y_i)$
at the lattice point $(x_i,y_i)$ of the physical two-dimensional space.  In
order for a notion of measure based on phase space volume to be useful, the
volume must be preserved in the course of time.

\subsubsection{Liouville's theorem}

Preservation of phase space volume is assured by Liouville's theorem, an
important theorem of mechanics.  As noted above, Burgers and Lee showed that a
version of Liouville's theorem applies to the system governing the Fourier
amplitudes for the inviscid fluid.  For vorticity dynamics the amplitudes
satisfy
\begin{equation}
\dot{q}_{\mathbf k} = \sum_{\mathbf{l},\mathbf{m}} \frac{\epsilon_{\mathbf{k}
\mathbf{l}\mathbf{m}}}{|{\bf l}|^2} (q_{\mathbf{l}} - h_{\mathbf l})\,
q_{\mathbf{m}}\,,
\end{equation}
where $h_{\mathbf l}$ is the Fourier transformation of the beta effect and
$\epsilon_{\mathbf{k}\mathbf{l}\mathbf{m}} = \hat{z} \cdot (\mathbf{l} \times
\mathbf{m}) \delta (\mathbf{k} + \mathbf{l} + \mathbf{m} )$ is completely
antisymmetric, i.e., $\epsilon_{\mathbf{k}\mathbf{l}\mathbf{m}} = -
\epsilon_{\mathbf{l} \mathbf{k}\mathbf{m}} = - \epsilon_{\mathbf{m} \mathbf{l}
\mathbf{k}}$ and $\epsilon_{\mathbf{k}\mathbf{k}\mathbf{m}} =
\epsilon_{\mathbf{k} \mathbf{l} \mathbf{k}} = 0$.  Therefore, antisymmetry
directly implies Liouville's theorem, $\sum_{\mathbf{k}}\partial \dot{q}
_{\mathbf{k}}/\partial q_{\mathbf{k}}\equiv 0 $.

Similarly, we have shown directly that the lattice model possesses a version of
Liouville's theorem, which we recently discovered was anticipated in
\cite*{burgers29b}.  This result was also inferred in \cite*{ST74}.  We assume
periodic boundary conditions.  The lattice model discretization can be viewed as
an expansion of the vorticity in terms of a tent function basis [e.g.\
\cite*{fletcher90}].  Upon multiplying the equation of motion by a basis
function, as is typical with Galerkin projection, and representing all
derivatives as differences, Eq.~(\ref{eq:euler}) becomes
\begin{equation}
\label{eq:JFM_Biuv}
\dot{q}_i =  \sum_{j,k} B_{ijk} \psi_{j} q_{k}\,,
\end{equation}
which is an equation for the potential vorticity at the lattice point $i$.
Assuming a periodic lattice, the quantity $B_{ijk}$ is easily seen to be
completely antisymmetric, i.e., $B_{ijk} = -B_{jik} = -B_{kji}$ and $B_{iji} =
B_{iik} =0$, just as was the case for
$\epsilon_{\mathbf{k}\mathbf{l}\mathbf{m}}$.  Therefore, Liouville's theorem
follows,
\begin{eqnarray}
\sum_i \frac{\partial \dot{q_i}}{\partial q_i} &=& \sum_{i,j,k} B_{ijk} (M_{ji}
q_{k} + \delta_{ki} \psi_j ) = \sum_{i,j} ( B'_{iij} q_j + B_{iji} \psi_j ) =
0\,,
\end{eqnarray}
where each term of the last sum vanishes. Here the matrix $M$ represents the
inverse Laplacian and $B'$ is another matrix that has the same antisymmetry
property as $B$.


\subsection{Canonical equilibrium distribution}

Having defined phase space and verified Liouville's theorem, we are poised to
write a partition function and to define phase space averages.  The natural
expression for the partition function associated with the canonical (Gibbs)
ensemble is
\begin{equation}
{\cal Z}_{c}=\int_{\cal G} e^{-\beta H[q] - C[q]} \,{\cal D}q\,,
\label{Zc}
\end{equation}
where $H$ is the Hamiltonian of \S\ref{sec:dynamics} and $C$ denotes the
infinite family of Casimir invariants.  Averages corresponding to (\ref{Zc}) are
given by
\begin{equation}
     \langle F \rangle_c
= \int_{\cal G} F[q] \, P_{c}[q;\beta,{\cal C}] \,{\cal D}q\,,
\label{<>c}
\end{equation}
where $F$ is a functional of $q$ and the phase space probability density is
given by
\begin{equation}
P_{c}[q;\beta,{\cal C}]= {\cal Z}_{c}^{-1}
e^{-\beta H[q] -C[q]}\,.
\label{gibbs}
\end{equation}
Expressions (\ref{Zc}) and (\ref{<>c}) are functional integrals
[\cite*{schulman,sundermeyer}], and the intent is to give them meaning by
discretizing as in \S\S\ref{ssec:state} and then taking the limit $N\rightarrow
\infty$ and $\delta\rightarrow 0$.  Finding unique well-defined results with
this procedure for such integrals, with other than quadratic functionals in the
exponent, is usually a difficult task.  Consequently, a mean field approach has
been taken, which we turn to in \S\ref{sec:MF}.

An alternative to the direct evalution of (\ref{<>c}) is to appeal to the fact
that the dynamics of (\ref{eq:euler}) is an area-preserving {\it rearrangement}
[e.g.\ \cite*{lieb01}].  This means for an initial condition $q_{0}$, the
solution at time $t$ is given formally by
$q(x,y,t)=q_{0}(x_{0}(x,y,t),y_{0}(x,y,t))$, where $(x_{0}(x,y,t),y_{0}(x,y,t))$
are the initial conditions of the characteristics, which satisfy
$\partial(x_{0},y_{0})/\partial(x,y)=1$.  The Casimir invariants are associated
with relabelling symmetry [e.g.\ \cite*{salmon82,padhye96}] and possess the same
value when evaluated on functions that are related by rearrangement.  Thus, if
one restricts the domain of integration $\cal G$ to be rearrangements of a given
function, denoted by ${\cal G}_{\cal R}$, then we should obtain the same answer
because $\langle F[q] \rangle_{\cal R} = F[q]$ for functionals with integrands
that depend only on $q$, such as Casimirs and $\exp({C[q]})$.  Here $\langle \
\rangle_{\cal R}$ is defined with $P_{\cal R}[q;\beta]= {\cal Z}_{\cal R}^{-1}
\exp{(-\beta H[q])}$ and ${\cal Z}_{\cal R}=\int_{{\cal G}_{\cal R}}
\exp{(-\beta H[q])} \,{\cal D}q$.

\section{Mean Field Approximation \& Statistical Independence}
\label{sec:MF}

It is well-known that vorticity equations like (\ref{eq:euler}), the Vlasov
equation, and other transport equations develop fine structure in the course of
time.  Because of this \cite*{lynden67} proposed a coarse graining procedure to
obtain a most probable state.  He divided phase space up into hyper-fine cells
that are assumed to be capable of resolving the fine structure.  These are the
$m$-cells referred to in \S\S\ref{ssec:state}, which have a scale size $\delta$.
Experimentally $\delta$ is determined by the resolution, but in ideal theory the
fine structure can become arbitrarily fine and so a limiting procedure is
required.  In addition \cite*{lynden67} proposed larger cells, which we have
called $M$-cells, that characterize a macroscopic scale $\Delta$.  The $M$-cells
contain many $m$-cells that can be freely exchanged within an $M$-cell without
changing any macroscopic quantity.  Thus one is able to count states and obtain
an expression for a coarse grained or mean field entropy that can be maximized
subject to constraints.  Later, \cite*{miller90} and \cite*{robert91}
reconstructed and improved this formulation.  Miller defined $m$-cells and
$M$-cells based on scales with the property that the energy averaged over
$M$-cells approximates the energy averaged over $m$-cells.  However, we argue
that the most important condition for separating the $M$-cell and $m$-cell scale
lengths is {\it statistical independence}, which assures near independence of
the probability densities of $M$-cells, which are viewed as subsystems, and is
associated with near additivity of the constraints.  These are crucial
properties.

\begin{figure} 
\begin{center}
\psfrag{Omega}{$\Omega$} \psfrag{q}{$\bar q$ (s$^{-1}$)} \psfrag{psi}{$\bar
\psi$ (cm$^2$/s)} \psfrag{qc}{$\bar q$} \psfrag{psic}{$\bar \psi$}
\psfrag{pi2}{$\pi/2$} \psfrag{pi}{$\pi$} \psfrag{3pi2}{$3\pi/2$}
\psfrag{2pi}{$2\pi$} \psfrag{a}{\bf (a)} \psfrag{b}{\bf (b)} \psfrag{c}{\bf
(c)} \psfrag{d}{\bf (d)} \psfrag{r(cm)}{$r$ (cm)} \psfrag{0}{0} \psfrag{1}{100}
\psfrag{2}{200} \psfrag{3}{300} \psfrag{4}{400} \psfrag{theta}{$\theta$}
\includegraphics[width=\linewidth]{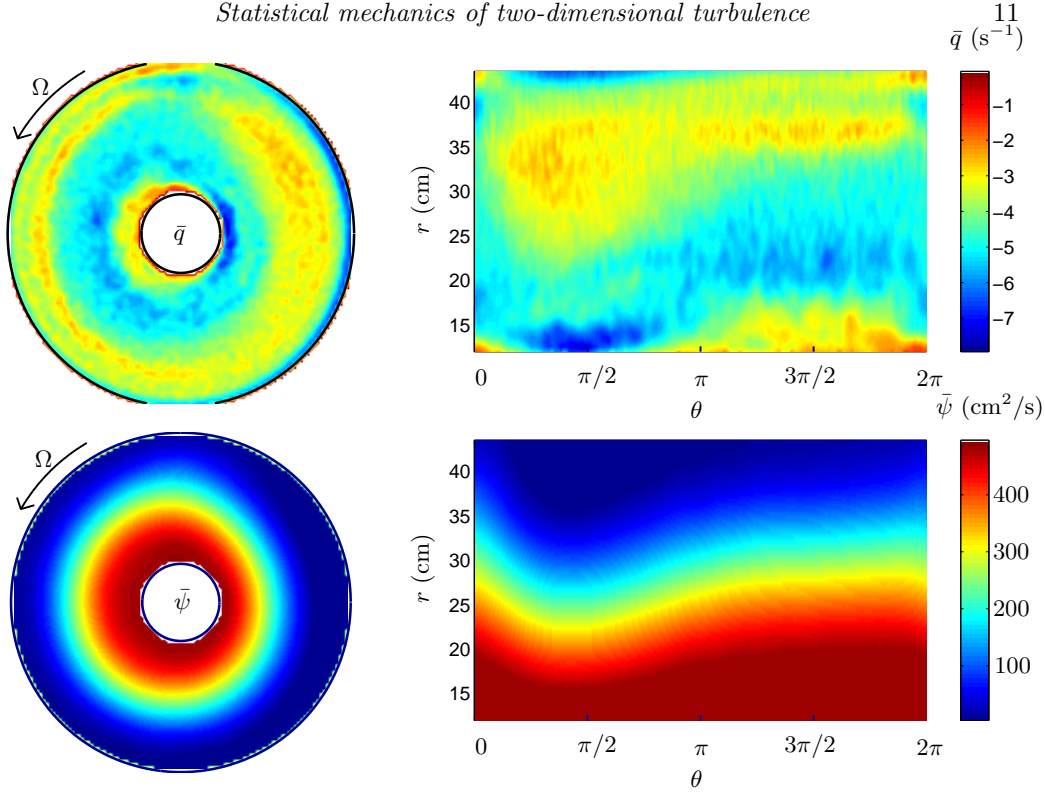}
\end{center}
\caption {The time-averaged potential vorticity (top two figures) and the
streamfunction (bottom two figures) in the Rossby wave frame. The figures on
the left show the fields in the rotating tank; the figures on the right show
the same fields unwrapped. The streamfunction field is smoother than the
potential vorticity field since the vorticity is given by a second derivative
(the Laplacian) of the streamfunction. Hence the characteristic length scale in
the azimuthal direction is larger for the streamfunction than for the potential
vorticity.} \label{fig:JFM_average}
\end{figure}

Experimentally the two scales can be demonstrated as in
Fig.~\ref{fig:JFM_average}.  Observe in the upper plot of this figure the fine
scale structure in the potential vorticity, while in the lower plot the
streamfunction, due to the integration over the Green's function, is
considerably smoother.  We take the upper scale to be $\delta$ and the lower
scale to be $\Delta$.

\subsection{Counting states}
\label{ssec:counting}


According to Lynden-Bell's statistics, the number of ways to distribute
$m$-cells into $M$-cells is
\begin{equation} \label{eq:JFM_count}
W = \prod_r \frac{N_r !}{\prod_I N^{(I)}_{r}!} \prod_I \frac{N^{(I)}
!}{\left(N^{(I)} - \sum_r N^{(I)}_{r}\right)!}
\end{equation}
where $N_r$ is the total number of $m$-cells with the $r$th value of potential
vorticity in the whole space, and $N^{(I)}_{r}$ is the total number of
$m$-cells with the $r$th value of potential vorticity in the $I$th $M$-cell.
Also, $N^{(I)}$ is the total number of m-cells in the $I$th $M$-cell.  The
first product in Eq.~ (\ref{eq:JFM_count}) represents the number of ways to
distribute $N_r$ $m$-cells into groups of $\{ N^{(I)}_{r} \}$, where $I$ counts
all $M$-cells and the second product is the number of ways to distribute inside
an $M$-cell.  Also, $N^{(I)}-\sum_r N^{(I)}_{r}$ can be understood as the
number of empty $m$-cells.  Lynden-Bell proposed this manner of counting for
stellar dynamics [\cite*{lynden67}, see also \cite*{ST132}], where $m$-cells
represent stars, which are considered to be distinguishable, and there may be
empty $m$-cells.  However, the statistics for the two-dimensional continuum
Euler model is a special case of Lynden-Bell's general counting procedure.

In Miller's application to the two-dimensional continuum Euler model, he
assumes that all $m$-cells are occupied by a vortex and these vortices are
indistinguishable if they have the same value of vorticity.  Because there are
no empty $m$-cells, $N^{(I)}=\sum_r N^{(I)}_{r}$ and because the $m$-cells are
indistinguishable, a factor of $\prod_r 1/(N_r!)$ is added.  This counting
produces
\begin{equation}
W = \prod_{I} \frac{N^{(I)}!}{\prod_r N_{r}^{(I)}!} \, .
\end{equation}
The above equation already involves statistical independence among different
$M$-cells.

Boltzmann articulated the entropy as a measure of the number of possible
configurations of the system.  Therefore, the entropy $S$ is defined to be the
logarithm of the total number of configurations, $\ln W $.  If $N_r^{(I)}$ is
large, Stirling's formula gives
\begin{equation} \label{eq:JFM_entropy}
S = {\ln W} \cong - \sum_{r,I} \left( {N_{r}^{(I)}} \right) \ln
\left(\frac{N_{r}^{(I)}}{N^{(I)}} \right).
\end{equation}
In the continuum limit of potential vorticity levels, $N_{r}^{(I)}/N^{(I)}$ is
replaced by $P_M(\zeta;x,y)$, and $\sum_{r,I}$ by $\int d\zeta dxdy$.  In short,
the index $I$ represents the coordinates for the discretized $M$-cells and the
index $r$ represents the ordered level sets of potential vorticity inside the
$M$-cells.  Thus, it is replaced by the continuum vorticity variable $\zeta$,
the vorticity on an $M$-cell.  With these observations, the resulting total mean
field entropy is seen to be
\begin{equation}
S_M [P_M] = - \int P_M(\zeta;x,y) \ln P_M(\zeta;x,y)\, d\zeta dxdy
= -\int  \langle \ln P_M \rangle_M \,dxdy
\label{Mentrpy}
\end{equation}
where $P_M(\zeta;x,y)$ is the probability density in the mean field
approximation. The density of $P_M(\zeta;x,y)$ is centered at the point $(x,y)$
and satisfies the normalization $\int P_M d\zeta = 1$. The integration over
$dxdy$ can be viewed as a sum over the $M$-cells that cover the domain of the
fluid. The second equality of (\ref{Mentrpy}) follows from the definition
$\langle A \rangle_M =\int A P_M d\zeta$, and thus $S_M [P_M]$ can be naturally
termed the (mean field) Boltzmann-Gibbs entropy.

In closing this subsection, we reiterate that the potential vorticity variable
$q$ is a field variable, a function of coordinates.  However, when we introduced
the probability density $P_M$ on $M$-cells, we used $\zeta$, an independent
variable, to represent the values of the potential vorticity on an $M$-cell.

\subsection{Mean field canonical distribution}
\label{ssec:meanCD}

Given the mean field entropy $S_M$ we can proceed to obtain the mean field
density $P_M(\zeta;x,y)$ as the most probable state by extremization subject to
particular mean field constraints.  These constraints and their corresponding
Langrange multipliers are given as follows:
\begin{enumerate}
\item The {\it Hamiltonian constraint} is obtained by replacing the vorticity
variable $q$ in $H[q]$ with its mean field average, to obtain a mean field
energy,
\begin{eqnarray}
  H_M [P_{M}] &=& \frac{1}{2}\int \left(\zeta  P_{M}(\zeta;x,y)- h\right)\,
  \zeta' P_{M}(\zeta';x',y')
G(x,y;x',y')\, d\zeta dxdy\, d\zeta'dx'dy' \nonumber \\
&=& \frac{1}{2}\int \langle \psi \rangle_{M}
\left(\langle \zeta \rangle_{M} - h\right) \, dxdy
\end{eqnarray}
where $\langle \zeta \rangle_{M}=\int \zeta P_{M} \,d\zeta$ and $\langle \psi
\rangle_{M}$ is defined by
\begin{equation}
     \langle \zeta
\rangle_{M}=-\nabla^{2}\langle \psi \rangle_{M} +h\,.
\label{Mpois}
\end{equation}
The Lagrange multiplier associated with this constraint is taken to be the
constant value, $-\beta$, where the minus sign is by convention.

\item The {\it normalization constraint} is $\int P_M d\zeta = 1$.  This is a
normalization on each $M$-cell; thus, although $P_M$ depends on
position, the integration does not. Because this is a constraint for each
point $(x,y)$, the Lagrange multiplier in this case depends on position.  We
call it $\gamma(x,y)$, and the quantity that appears in the variational
principle is
\begin{equation}
  N_{M}[P_{M}]=\int \gamma(x,y) P_M(\zeta;x,y) \, d\zeta dxdy\,.
\end{equation}

\item The {\it mean field Casimir constraint}, roughly speaking, contains the
information that on average, the area between any two contours of vorticity
remains constant in time.  More precisely, the quantity $g(\zeta)=\int P_{M}\,
dxdy$ is taken to be constant.  Because this is true for all $\zeta$, the
Lagrange multipler $\mu$ is likewise a function of $\zeta$ and the constraint
can be written as
\begin{equation}
  C_{M}[P_{M}]=-\beta\int \mu(\zeta) g(\zeta)\,  d\zeta
  = -\beta\int \mu(\zeta) P_M(\zeta;x,y) \, d\zeta dxdy\,,
\end{equation}
where the prefactor of $-\beta$ is again by convention.  This constraint is the
mean field version of the family of Casimir invariants $C[q]$.
\end{enumerate}

Now we are in position to obtain the most probable state by extremizing the
quantity $F_{M}= S_{M}-\beta H_{M} + N_{M} + C_{M}$, i.e.,  upon functional
differentiation with respect to $P_{M}$, $\delta F_{M}/\delta
P_{M}=0$   implies
\begin{equation}
P_M(\zeta;x,y;\beta, \mu) = {\cal Z}_M^{-1}
e^{-\beta \left[\zeta\langle \psi \rangle_{M} - \mu(\zeta)\right]}\,,
\label{Mgibbs}
\end{equation}
where ${\cal Z}_M = \int e^{-\beta [\zeta\langle \psi \rangle_{M} - \mu(\zeta)]}
d\zeta$ and evidently $P_M$ is normalized.  Equation (\ref{Mgibbs}) is the mean
field counterpart to (\ref{gibbs}) and could aptly be termed the canonical
(Gibbs) mean field distribution.  The above variational principle and extremal
distribution (\ref{Mgibbs}) appeared in essence in an appendix of
\cite*{lynden67}.

Given (\ref{Mgibbs}) we are in a position to calculate $\langle
\zeta\rangle_{M}$ and then substitute the result into (\ref{Mpois}).  This gives
the mean field Poisson equation,
\begin{equation}
\nabla^{2}\langle \psi \rangle_{M}= {\cal Z}_M^{-1} \int \zeta \, e^{-\beta
\left[\zeta\langle \psi \rangle_{M} - \mu(\zeta)\right]}\,  d\zeta  + h\,.
\label{Mpois2}
\end{equation}
Versions of this equation have been solved in various references [e.g.\
\cite*{robert91,miller90,ST112}], but we will not do this here.


We conclude this subsection by giving a heuristic connection between $\langle \
\rangle_M$, a prescription for averaging  functions, and $\langle \ \rangle_c$,
a prescription for averaging  functionals.  Consider the functional $q(x',y')$,
by which we mean the evaluation of the function $q$ at the point $(x',y')$, and
evaluate
\begin{equation}
     \langle q(x',y')\rangle_c
= \int_{\cal G} q(x',y') \, P_{c}[q;\beta,{\cal C}] \,{\cal D}q\, .
\label{<z>c}
\end{equation}
If we rewrite (\ref{<z>c}) as an integral on $M$-cells, where $q(x',y')$ is
$q_{I'}$, write ${\cal D}q = \prod_{J} dq_{J}$, and then assume statistical
independence of $M$-cells, $P_{c}=\prod_{I} P_{I}$, we obtain
\begin{equation}
\langle q(x',y')\rangle_c = \int  q_{I'}\, \prod_{I} P_{I} \,\prod_{J} dq_{J}
=\int  q_{I'}\,  P_{I'} \,  dq_{I'} = \int \zeta P_{M} d\zeta =  \langle \zeta
\rangle_M\,. \label{<q>c}
\end{equation}  This derivation emphasizes the need for near statistical
independence of $M$-cell subsystems.


\subsection{Ruggedness \& additivity}
\label{ssec:rugadd}

Classical statistical mechanical treatments of the canonical ensemble allow for
subsystems to interact and exchange energy, but their interaction is assumed to
be weak and the details of the interaction are usually ignored in calculations.
Neglect of the interaction energy results in the energy being equal to the sum
of the energies of the individual subsystems, i.e., the energy is an additive
quantity.  In conventional treatments only additive invariants are used in
calculating the most probable distributions, and in some treatments [e.g.\
\cite*{landau_ST}] this requirement is explicitly stated.  The reason for this
is that additive invariants give rise to statistical independence of
subsystems. In our treatment of fluids, subsystems are $M$-cells and so we
consider invariants that are additive over these regions.  There is a close
connection between ruggedness of invariants and the property of additivty.  We
show that only the rugged invariants are additive, and thus they characterize
the statistical properties of $M$-cells.  In \S\S\ref{ssec:sub} and
\S\S\ref{ssec:pdf} we will see that experimental results support this
reasoning.

\cite*{kraichnan80} Fourier transformed and truncated to obtain a
finite-dimensional system.  They argued that the truncated remnants of the
total vorticity, enstrophy, and energy are the only invariants to be used in a
statistical mechanics treatment because these invariants are rugged, i.e.,  they
remain invariants of the truncated system.  They  also appear to be aware that
these invariants possess the property of additivity, but they do not emphasize
this point.  Although \cite*{ST74} has argued that this kind of truncation does
not properly handle small scale behavior, we find that this theory does a
fairly good job at predicting the energy spectrum, but we will report on this
elsewhere.  We argue in general that such invariants are important because they
are the only additive invariants. Below we consider a somewhat more general
setting.

Because of Parseval's identity, the quadratic invariants are additive
and higher order invariants are not.  To see this, suppose we define
$M$-cells to be composed of amplitudes of some subsets of Fourier
modes, which we denote by $\kappa_{I}$.  Then a sum over modes can be
done in groupings, i.e.\ $\sum_{\mathbf k}=\sum_{I}\sum_{\kappa_{I}}$.
(This is the idea behind spectral reduction [\cite*{bowman99}], a
computational method where groupings of Fourier modes (bins) are
described by a single representative\@.)  For the quadratic Casimir
invariant, the enstrophy, we have
\begin{equation}
C_{2}=\int q^{2}\, dxdy =(2\pi)^{2}\sum_{\mathbf{k}}
|q_{\mathbf{k}}|^{2}\,,
\end{equation}
and defining an $M$-cell enstrophy by $C_{2}^{(I)}=(2\pi)^{2}\sum_{\kappa_{I}}
|q_{\mathbf{k}}|^{2}$, we obtain $ C_{2} = \sum_{I}
C_{2}^{(I)}$. Similarly, the energy can be written as a sum over
$M$-cell energies, $E = \sum_{I} E^{(I)}$.  The linear Casimir
invariant $C_{1}=\int q dxdy$ merely reduces to the zeroth
Fourier coefficient, and is thus in a trivial sense additive. Higher
order invariants,  $C_{n}=\int q^{n} dxdy$ for $n>2$,  have Fourier
representations that are not reducible to expressions in terms of a
single sum over $M$-cells.

The discretized lattice model has properties similar to those described above.
The quadratic Casimir invariant and energy reduce to sums over a finite number
of $m$-cell lattice variables, $q_i$, $h_i$ and $\psi_i$, which are potential
vorticity, height, and streamfunction represented in terms of the kernel
function $K_i$ of \S\S \ref{sec:SM} as follows:
\begin{eqnarray}
C_2 &=& \int q^2 \, dxdy = \sum_{i,j} \int K_i K_j q_i q_j \,dxdy
= \sum_{i,j} q_i {Z}_{ij} q_j \, , \nonumber
\\
H &=& \frac1{2}\int q \psi dxdy = \frac1{2} \sum_{i,k} \int K_i K_k (q_i-h_i)
\psi_k\, dxdy  \nonumber \\
&=& \frac1{2} \sum_{i,k} (q_i-h_i) {Z}_{ik} \psi_k = \sum_{i,j} (q_i-h_i)
\hat{Z}_{ij} (q_j-h_j)
\end{eqnarray}
where ${Z}_{ij} = \int K_i K_j \, dxdy$ and $\hat{Z}_{ij} =\sum_{k} {Z}_{ik} M_{kj}$ are
symmetric commuting matrices.  These invariants are rugged, i.e, they are
conserved by the finite dynamical system obtained by projection onto the
lattice.  In addition, because ${Z}$ and $\hat{Z}$ commute, one can
always find an orthogonal matrix ${\cal O}$ that satisfies ${Z} = {\cal
O}^T\!\!D{\cal O}$ and $\hat{Z} = {\cal O}^T\!\!\hat{D} {\cal O}$, where
$D_{ij}= d_i \delta_{ij}$ and $\hat{D}_{ij}= \hat{d}_i \delta_{ij}$ are diagonal
matrices.  Defining $q' = q{\cal O}$, $h' = h{\cal O}$ and $\psi' = \psi {\cal
O}$, the enstrophy and energy become
\begin{eqnarray}
C_2 &=& \sum_{i,j} q'_i D_{ij} q'_j =  \sum_i d_i (q'_i)^2 =
\sum_{I}\sum_{\kappa_I} d_i (q'_i)^2 \, \nonumber
\\
H &=& \sum_i \hat{d}_i (q'_i - h'_i)^2 = \sum_{I}\sum_{\kappa_I} \hat{d}_i
(q'_i - h'_i)^2 \, ,
\end{eqnarray}
where $I$ is the index for the $I$th $M$-cell and $\kappa_I$ denotes the set of
$m$-cells in the $I$th $M$-cell.

This coordinate transformation simultaneously diagonalizes the quadratic
Casimir invariant and the energy. However, higher-order Casimir invariants are
in general not rugged and are in general not simultaneously diagonalizable.
Thus, higher order invariants are not additive, which means $M$-cells  share
contributions from these invariants.  In this sense, invariants of order higher
than quadratic are not useful for describing the statistics of $M$-cells, which
by assumption are independent.

\subsection{Statistically independent subsystems}
\label{ssec:sub}

Now we turn to the question of how to find subsystems, i.e, how to a find a
good definition of the $M$-cells.  First we note that flows inside the rotating
tank with the sloped bottom have azimuthal undulations in most physical
quantities (streamfunction, potential vorticity, etc.), and these undulations
have been identified as Rossby waves [\cite*{DCN92,solomon93}]. In a co-rotating
frame, these waves propagate in the rotation direction at constant velocity.
Thus, by shifting to a frame moving at the phase velocity of the Rossby wave,
we obtain a pattern that is statistically stationary on large scales.  For
example, the wavy patterns corresponding to the time-averaged streamfunction
and potential vorticity are shown in Fig.~\ref{fig:JFM_average}.  As noted
before, the streamfunction is fairly smooth, characteristic of the scale
$\Delta$, is monotonically decreasing in the radial direction, and describes a
strong zonal flow.  However, the time-averaged potential vorticity is scattered
with fine structure in space, the $\delta$ scale, but still has a wavy mean
pattern similar to that of the time-averaged streamfunction.   So, this suggests
that the first step toward defining $M$-cells is to consider a frame moving at
the phase velocity of the Rossby wave.

\begin{figure}
\begin{center} 
\psfrag{radial}{$\Delta r$ (cm)} \psfrag{theta}{$r \Delta \theta$ (cm)}
\psfrag{0.75}{0.75} \psfrag{0.80}{0.80} \psfrag{0.85}{0.85} \psfrag{0.90}{0.90}
\psfrag{0.95}{0.95} \psfrag{0}{0} \psfrag{-2}{-2} \psfrag{2}{2} \psfrag{-4}{-4}
\psfrag{4}{4} \psfrag{-6}{-6} \psfrag{6}{6} \psfrag{0 }{0} \psfrag{-2 }{-2}
\psfrag{2 }{2}
\includegraphics[width=\linewidth]{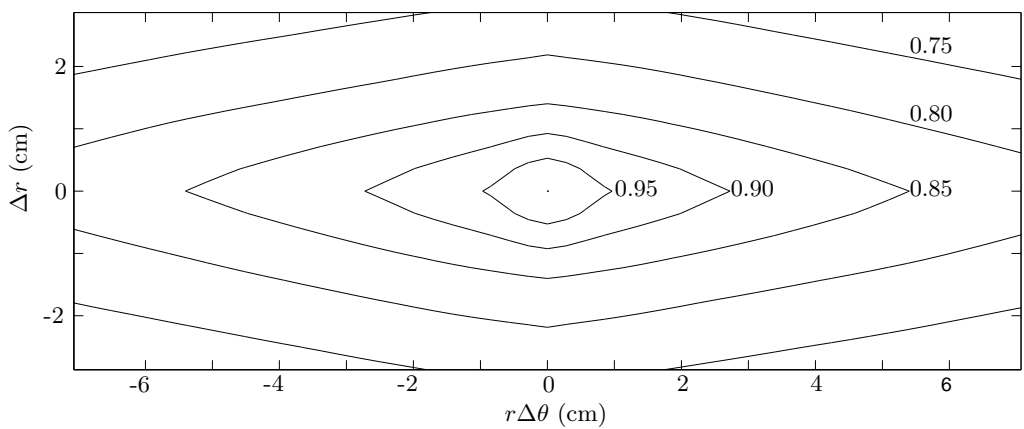}
\end{center}
\caption{Contours of the correlation function $C_{\mathrm{cor}}(\Delta r,
r \Delta \theta)$ illustrating the anisotropic nature of the potential vorticity field,  which has longer range correlation in the azimuathal direction than in the radial
direction (cf. Fig.~\ref{fig:JFM_average}).} \label{fig:JFM_corr}
\end{figure}

Having determined the frame, we seek $M$-cells that are statistically
independent.  Because strong correlation in a preferential direction might
affect the geometry of $M$-cells and associated additive invariants, we have
measured the correlation function,
\begin{equation}
C_{\mathrm{cor}}(\Delta r, r \Delta \theta) = \frac1{T}\int_{0}^{T}
\frac{\int q(r,r\theta;t) q (r + \Delta r ,r \theta + r\Delta \theta ;t) r dr
d\theta}{\int q (r,r \theta;t)^2 r dr d\theta} dt \,, \label{cor}
\end{equation}
where $(\theta,r)$ are the usual polar coordinates. From a large data set of
PIV measurements we obtain the time average of the velocity field, whence we
calculate the potential vorticity at different positions.  Then the integrals
of (\ref{cor}) are performed with the spatial limits being the bulk of the area
occupied by the fluid with a resolution of $\delta \approx 0.8\mathrm{~cm}$ and
the time limit taken to be 80 revolutions with 47 measurements.  The result of
this procedure is presented in Fig.~\ref{fig:JFM_corr}, which shows contours of
$C_{\mathrm{cor}}$ plotted on a $\Delta \theta-\Delta r$ plane.  The highly
anisotropic nature of the contours suggests there is significantly less
correlation in the radial direction than in the azimuthal direction. Thus to
achieve consistent independence the shape of an $M$-cell should be elongated.

In the course of tracking blobs of fluid we generally observe that to good
approximation such blobs follow contours of the time-averaged streamfunction.
This, together with the the $C_{\mathrm{cor}}$ plot, suggests that a good
coordinate for dividing the system into subsystems is the time-averaged
streamfunction,
\begin{equation}
  \bar{\psi}(r,\theta) =\frac1{T}\int_{0}^{T} \psi(\theta,r;t) \,dt\,.
\end{equation}
Contours of $\bar \psi$ tend to be smooth and, we argue, are part of a natural
coordinate system for describing turbulence with a mean flow that has slow
spatial dependence.  (We have also considered $\bar q$ but found it to be not
as good because of its greater variability.)  To complete the coordinate
system, we introduce a coordinate $\chi$, which is conjugate to $\bar{\psi}$
and therefore satisfies
\begin{equation}
\frac1{r}  \frac{\partial \bar{\psi} }{\partial \theta}\frac{\partial
\chi}{\partial r} - \frac1{r} \frac{\partial \bar{\psi} }{\partial r}
\frac{\partial \chi }{\partial \theta} = 1\,.
\end{equation}
Thus the coordinate transformation $(\theta,r)\longleftrightarrow
(\chi,\bar\psi)$
satisfies $rdrd\theta=d\chi d\theta$.

We propose that contours of $\bar \psi$ define $M$-cells, which we take to be of
small (infinitesimal) width in this coordinate, and we propose that the $\chi$
coordinate at fixed $\bar\psi$ represents a continuum of $m$-cells.  We imagine
an $M$-cell to be a region (nearly a curve) at fixed $\bar\psi$.  Hence with
this definition, the probability density $P_{M_{exp}}$, depends only on the
potential vorticity variable $\zeta$ and on the coordinate $\bar\psi$; i.e.,
$P_{M_{exp}}(\zeta; \bar\psi)$ is the probability of finding a potential
vorticity value $\zeta$ in the $\bar \psi$ $M$-cell.  Thus the ensemble average
of an arbitrary function $f$ is written as
\begin{equation}
\label{eq:JFM_f_odot}
\langle f \rangle_{M_{exp}} (\bar \psi) = \int f(\zeta,
\bar \psi) P_{M_{exp}} (\zeta; \bar{\psi}) \,d\zeta\,, \label{podot}
\end{equation}
where $P_{M_{exp}}$ is normalized as $\int P_{M_{exp}} d\zeta = 1$.   In
practice we can determine the probability $P_{M_{exp}}$ from data by the
relative frequency definition (cf.~\S\S \ref{ssec:pdf}),  and then proceed to
calculate (\ref{podot}).  However, this is equivalent to averaging over $\chi$
and $t$; e.g.\ $\langle \zeta \rangle_{M_{exp}} = \overline{\prec \! q \!
\succ_{\chi}}$, where $\prec \! q \! \succ_{\chi}=\int q \,  d\chi/\int d\chi$.
Given $\langle \zeta \rangle_{M_{exp}}$  and using   (\ref{Mpois}) to define
$\langle \psi \rangle_{M_{exp}}$ we similarly have the  equivalence $\langle
\psi \rangle_{M_{exp}} (\bar \psi) =  \overline{\prec \! \psi (\bar \psi,
\chi;t) \! \succ_{\chi}}= \bar{\psi}$, where the second equality follows by
definition. The undular streamfunction of Fig.~\ref{fig:JFM_average} mainly
represents Rossby waves.  These wavy patterns are quite robust and often behave
as barriers to mixing.  In the Rossby wave frame, our data indicate that the
instantaneous streamfunction is close to the time-averaged streamfunction, i.e., 
$\prec \! \psi (\bar \psi, \chi;t)  \! \succ_{\chi}$ deviates from $\bar \psi$
by less than 10 percent. The  above comments can be viewed as  an experimental
verification of  ergodicity.

\begin{figure} 
\psfrag{psicms}{$\bar \psi$ (cm$^2$/s)} \psfrag{DeltaTC}{$\Delta_T\!C_2$}
\psfrag{DeltaTH}{$\Delta_T\!H$} \psfrag{DeltaC}{$\Delta_{\Psi}\!C_2$}
\psfrag{DeltaH}{$\Delta_{\Psi}\!H$} \psfrag{time}{time (s)} \psfrag{(a)}{\bf
(a)} \psfrag{(b)}{\bf (b)} \psfrag{(c)}{\bf (c)}
\includegraphics[width = \linewidth]{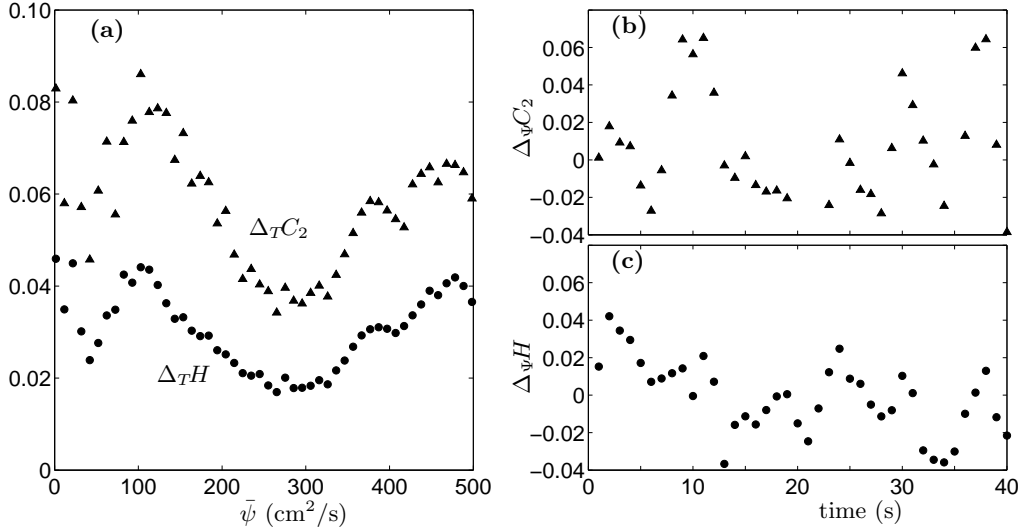}
\caption{(a) Enstrophy fluctuations $\Delta_T\!C_2(\bar\psi)$ [Eq. (\ref{CT})]
and energy fluctuations $\Delta_T\!H(\bar\psi)$ as a function of $\bar \psi$.
The fluctuations are small, indicating that energy and enstrophy are nearly
conserved for our choice of subsystem.  (b) Total enstrophy variations
$\Delta_{\Psi}\!C_2(t)$ [Eq. (\ref{CX})] and (c) total energy variations
$\Delta_{\Psi}\!H_2(t)$ with time; the variation is small, indicating that the
quantities for our choice of subsystem are almost conserved in time.}
\label{fig:fluct_energy_enstrophy}
\end{figure}

In terms of the above notation the  energy and enstrophy densities on
$M$-cells can be written as
\begin{eqnarray}
\langle H \rangle_{M_{exp}} (\bar{\psi}) &=& \frac{1}{2} \left[ \int \zeta
\bar{\psi} P_{M_{exp}} (\zeta;\bar{\psi}) \,d \zeta -
\prec \! \bar\psi h(\chi,\bar\psi) \! \succ_{\chi}
\right] \nonumber\\
&=& \frac{1}{2} \overline{\prec \! \bar\psi [q(\chi,\bar\psi,t)
-h(\chi,\bar\psi)]\! \succ_{\chi}}\,,
\nonumber \\
\langle C_2 \rangle_{M_{exp}} (\bar{\psi}) &=& \frac{1}{2} \int \zeta^2
P_{M_{exp}} (\zeta;\bar{\psi}) \,d \zeta =  \frac{1}{2} \overline{\prec \!
q^2(\chi,\bar\psi,t )\! \succ_{\chi}} \,,
\end{eqnarray}
and two quantities that measure spatial and temporal fluctuations of these
invariants can be compactly written as follows:
\begin{eqnarray}
\Delta_T\!C_2(\bar\psi) &=& \frac{ \left[ \, \overline{\left( \prec \! q^2
\! \succ_{\chi} - \overline{\prec \! q^2 \! \succ_{\chi}} \right)^2 } ~ \right]^{1/2} }{
\overline{\prec \! q^2 \! \succ_{\chi}} }\,,
\label{CT} \\
\Delta_{\Psi}\!C_2(t) &=& \frac{ \left[ \prec \! \left( \prec \! q^2 \!
\succ_{\chi}   - \prec \! q^2 \! \succ_{\chi\bar\psi} \right)^2 \!
\succ_{\bar\psi} \right]^{1/2} }{ \prec \! q^2 \! \succ_{\chi\bar\psi}
}\label{CX}\,,
\end{eqnarray}
with  similar expressions for $\Delta_T\!H(\bar\psi)$ and
$\Delta_{\Psi}\!H(t)$. Figure~\ref{fig:fluct_energy_enstrophy} depicts these
quantities. Panel (a) shows temporal fluctuations as a function of the spatial
coordinate $\bar \psi$. The middle regions of the experiment, where strong
zonal flows exist, is describable by  statistical mechanics. However, near the
walls, corresponding to high  and low $\bar \psi$ values, statistical mechanics
fails because of large fluctuations. Similarly, in panel (b) the spatial
fluctuations are plotted versus time, and it is observed that these
fluctuations are quite small. We have measured similar quantities for the cubic
and quartic Casimir invariants and the fluctuations are two or three times
greater.

An integrated measure of the goodness of our streamfunction based $M$-cells is
displayed in Table~\ref{tab:JFM}.  Here we have integrated $\prec \! \Delta_T H
\! \succ_{\bar \psi}$ and $\prec \! \Delta_T C_{2} \! \succ_{\bar \psi}$ over central
values of $\bar \psi$ and compared them with counterparts derived using square
cells.  By this measure streamfunction based cells are nearly ten times better
than square cells.

\begin{table}

\begin{center}
\begin{tabular}{|c||c|c|}
\hline Fluctuation Measure& $\prec \! \Delta_T H \! \succ_{\bar \psi}$
& $\prec \! \Delta_T C_{2} \! \succ_{\bar \psi}$ \\
\hline \hline
square cells  & 0.2233 & 0.6425 \\
\hline 
streamfunction cells  & 0.0343 & 0.0627 \\
\hline
\end{tabular}
\caption{Comparison of fluctuations for square cells with our streamfunction
based cells.  Both the energy fluctuation measure $\prec \! \Delta_T H \!
\succ_{\bar \psi}$ and enstrophy fluctuation measure $\prec \! \Delta_T C_{2}
\! \succ_{\bar \psi}$ are considerably smaller with the streamfunction based
cells.  These small fluctuations allow the division of the system into
$M$-cells, consistent with the statistical independence and additivity
assumptions of statistical mechanics.} \label{tab:JFM}
\end{center}
\end{table}

Thus, in summary, we have strong evidence supporting the use of streamfunction
based $M$-cells.  The evidence of Fig.~\ref{fig:fluct_energy_enstrophy} and
Table~\ref{tab:JFM} imply both statistical independence and the additive nature
of the quadratic invariants of these macro-cells.

\subsection{Prediction for PDFs}
\label{ssec:pdf}

\begin{figure}
\begin{center} 
\psfrag{Po}{$\!P_{M_{exp}}$} \psfrag{Pt}{$P^{tot}$}
\psfrag{Gaussian}{Gaussian} \psfrag{zeta/zetarms}{$(\zeta-\langle \zeta
\rangle_{M_{exp}})/{\zeta_{rms}}$} \psfrag{(a)}{\bf (a)} \psfrag{(b)}{\bf
(b)} \psfrag{(c)}{\bf (c)} \psfrag{(d)}{\bf (d)} \psfrag{Linear}{\bf Linear}
\psfrag{Logarithm}{\bf Logarithm}
\includegraphics[width=\linewidth]{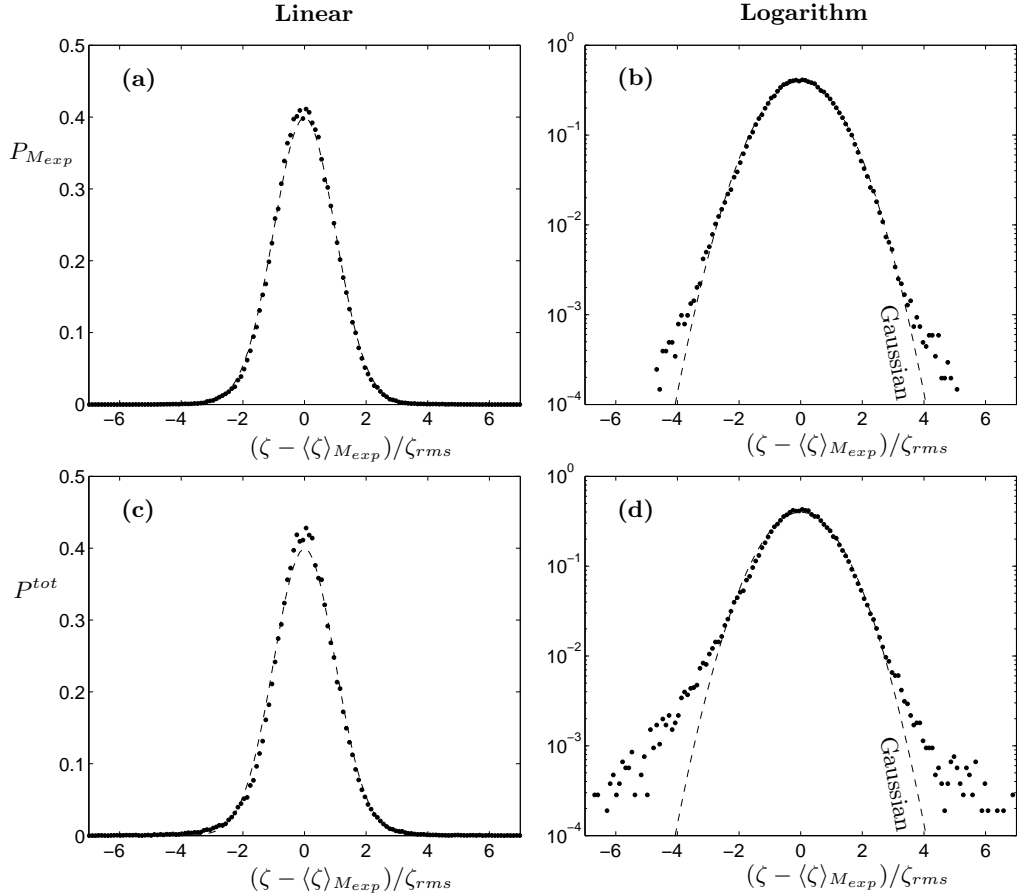}
\end{center}
\caption{The measured probability distribution of potential vorticity (data
points) on a typical $M$-cell is nearly Gaussian (dashed line), in accord with
(\ref{eq:JFM_P_odot}), as illustrated by these plots on (a) linear and (b)
logarithmic scales. In contrast, the potential vorticity of the whole system,
shown in (c) and (d) respectively, departs significantly from a Gaussian.}
\label{fig:stat_gauss}
\end{figure}

Based on the arguments in the previous section, we consider only two invariants
out of the infinitely many invariants conserved by the ideal dynamics.
Consequently, we obtain the following equilibrium distribution:
\begin{equation}
\label{eq:JFM_P_odot} P_{M_{exp}}(\zeta;\bar \psi)  = {\cal Z}^{-1}_{M_{exp}}
e^{-\beta \bar \psi \zeta - \gamma \zeta^2} \, ,
\end{equation}
where ${\cal Z}_{M_{exp}}=\int e^{-\beta \bar \psi \zeta - \gamma
\zeta^2} d\zeta$ depends only on $\bar \psi$.  Note, the function $h$
has cancelled out in the normalization.  This probability density
function (PDF) has the form of Gaussian that is shifted by
$\beta\bar{\psi}/2 \gamma$, a position dependent term that can be
interpreted as a sort of `local wind'.

In Fig.~\ref{fig:stat_gauss} we compare (\ref{eq:JFM_P_odot}) with experimental
results.  Figures~\ref{fig:stat_gauss}(a) and \ref{fig:stat_gauss}(b) show that
experimental data on
a typical  $M$-cell closely agree with the Gaussian distribution of
(\ref{eq:JFM_P_odot}).  
Each distribution is shifted by its mean value of potential vorticity $\langle
\zeta \rangle_{M_{exp}}$.  Figures~\ref{fig:stat_gauss}(c) and
\ref{fig:stat_gauss}(d) show the total probability $P^{tot}(\zeta)$, which is
the sum of the probabilities over all the $M$-cells, i.e., $P^{tot}(\zeta) =
\prec \! P_{M_{exp}}(\zeta;\bar \psi) \, \! \succ_{\bar \psi}$. 
These plots are decidedly non-Gaussian.

\begin{figure}
\begin{center} 
\psfrag{<zeta>}{\hspace{-5mm} $\langle q \rangle_{M_{exp}}$ (s$^{-1}$)}
\psfrag{psi}{$\bar \psi $ (cm$^{2}$/s)}
\includegraphics[width=.8\linewidth]
{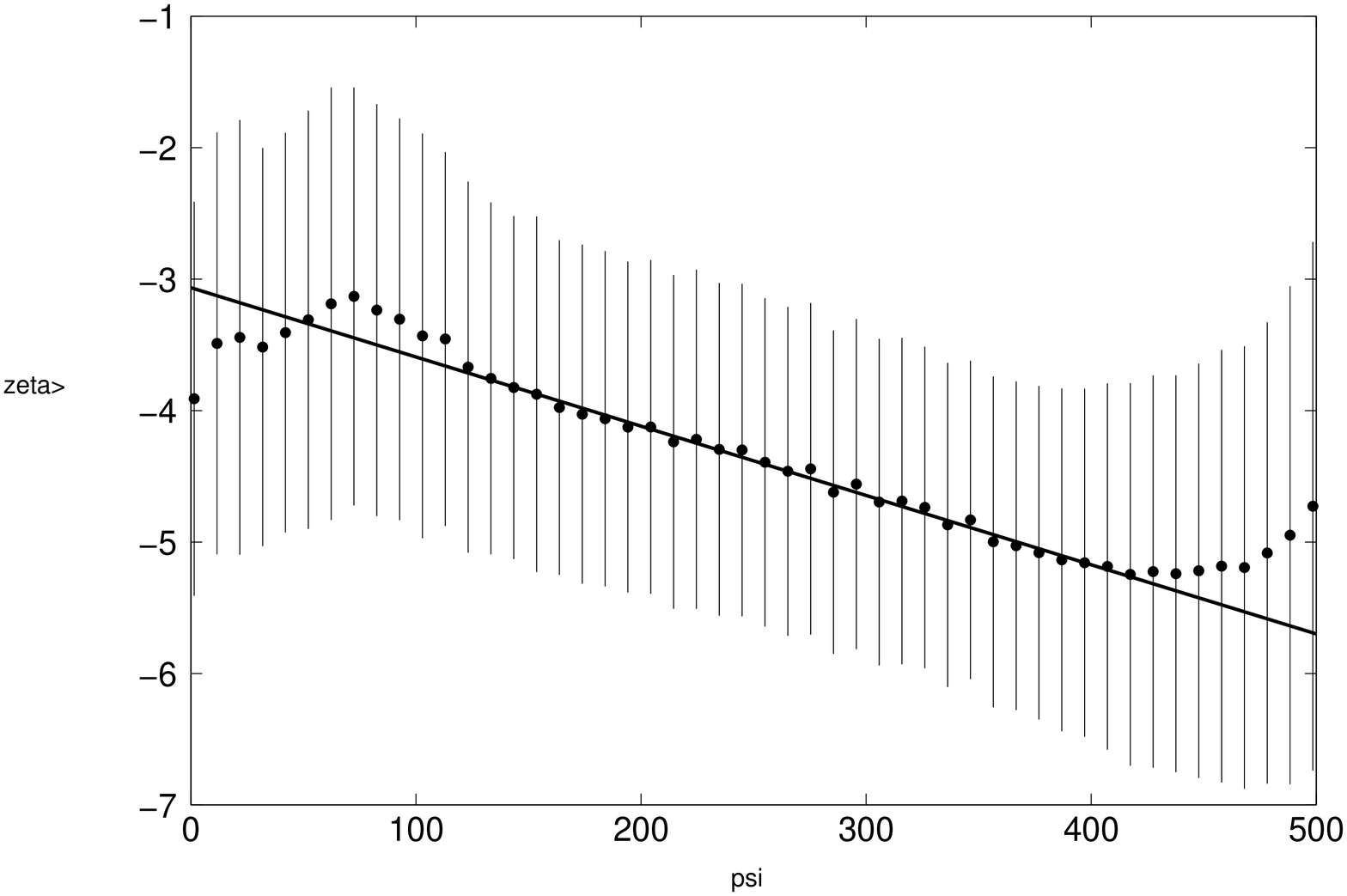}
\end{center}
\caption{The ensemble-averaged potential vorticity $\langle q
\rangle_{M_{exp}}$ exhibits a dependence on the time-averaged streamfunction
$\bar \psi$ that is linear except near the walls (at the ends of the range of
$\bar \psi$). The dots are mean values of $\langle q \rangle_{M_{exp}}$, and
the vertical lines correspond to standard deviations of $\langle q
\rangle_{M_{exp}}$ at a fixed $\bar \psi$. The data fit well the straight line
(a least-squares fit), in accord with the prediction of
Eq.~(\ref{eq:stat_linear}), where the slope is the ratio of two Lagrange
multipliers.} \label{fig:linear}
\end{figure}


The source of the small deviation in the tails of the distribution
plotted in Fig.~\ref{fig:stat_gauss}(b) can be attributed to
experimental resolution and sample size.  We analyzed this by
considering the scaling of the skewness and kurtosis with respect to
the number of data points $N$ and the experimental subsystem width
$\Delta\bar\psi$.  For a gaussian distribution the kurtosis is three
and the skewness is zero.  If we fix $N$ (large), our data indicate
that the kurtosis approaches three as $(\Delta\bar\psi)^a$, where
the exponent $a$ is less than unity and is approximately $0.5$
for our last (smallest $\Delta \bar\psi$) data points.  At fixed
$\Delta\bar\psi$ (small) we find that the tails decrease as we
increase $N$.  A similar examination of the skewness reveals
randomness about the value zero.


The next question is, what is the most probable value of potential vorticity in
each $M$-cell?  The probability distribution of Eq.~(\ref{eq:JFM_P_odot}) gives
a relation between the averaged vorticity and the streamfunction,
\begin{eqnarray}
\langle {\zeta} \rangle_{M_{exp}} &=& \int \zeta
P_{M_{exp}}(\zeta;\bar{\psi})\, d\zeta = - \epsilon \bar{\psi}\,,
\label{eq:stat_linear}
\end{eqnarray}
which follows by elementary integration.  Here $\epsilon=\beta/(2\gamma)$ is
the ratio of two Lagrange multipliers.  Figure~\ref{fig:linear} shows a linear
relation between the ensemble-averaged potential vorticity $\langle {\zeta}
\rangle_{M_{exp}}$ and the time-averaged streamfunction $\bar \psi$, as
predicted by Eq.~(\ref{eq:stat_linear}).

Therefore, our theoretical predictions based on a mean field approximation are
in good accord with PDFs on $M$-cells and the averaged values of potential
vorticity and streamfunction from experiments.  Our theory also indicates that
equilibrium can be locally achieved in $M$-cells, even though the system as a
whole is turbulent and non-Gaussian.

\section{Conclusions}
\label{sec:discuss}

In this paper we have emphasized the relationship between additive
invariants and statistical independence: probability densities that
result from entropy maximization principles, such as that of
\S\S\ref{ssec:meanCD}, will decompose into a product over subsystems
if the entropy is logarithmic (extensive) and the invariants included
as constraints are
additive over subsystems (M-cells).  We have also emphasized that
additivity and, consequently, independence depend on the definition of
subsystem.  This idea appears, at least implicitly, in conventional
statistical mechanics.  For example, in the classical calculation of
the specific heat of a solid, where one considers a solid to be a
collection of lattice sites with spring-like nearest neighbor
interactions, the Hamiltonian achieves the form of a sum over simple
harmonic oscillators.  However, such a diagonal form requires the use
of normal coordinates, and only then is the partition function equal
to a product over those of the individual oscillators.  Thus the
notions of subsystem (here a single oscillator), additivity, and
statistical independence are intimately related.

In our application of statistical mechanics to inhomogeneous damped
and driven turbulence, we have discovered experimentally that a good
definition of subsystem is provided by the temporal mean of the
streamfunction.  With this definition, the quadratic invariants
(energy and enstrophy) are additive, and the concomitant probability
density of (\ref{eq:JFM_P_odot}) agrees quite well with experimental
results for both the distribution of vorticity, as depicted in
Fig.~\ref{fig:stat_gauss} (a) and (b), and the mean state, as depicted
in Fig.~\ref{fig:linear}.

An alternative interpretation of our results can be obtained by the counting
argument of \S\S\ref{ssec:counting}.  Our definition of subsystem amounts to
the idea that potential vortices on the same contour of time-averaged
streamfunction can exchange their positions with little change in the energy
and enstrophy. However, the relocation of two potential vortices that are on
different contours of the streamfunction should result in a large change of the
invariants. In this sense, the number of possible configurations in phase space
can be counted, and the maximization of the entropy so obtained gives our
result.


Our discussion of statistical independence and additivity has been
heuristic, in the spirit of Boltzmann and Gibbs.  We suggest that a
more rigorous development could use the techniques described in other
works [e.g.\ \cite*{miller92,ST112,ST74}] adapted to our $\bar\psi$
coordinate that describes our subsystems.  For example, one  could begin
with an appropriate sequence of lattice models and obtain a continuum
limit.


Although in this paper we have focused on a geostrophic fluid, our procedure is
of general utility and is applicable to physical systems governed by a variety
of transport equations.  The unifying formalism is the noncanonical Hamiltonian
description of \S\ref{sec:dynamics}, which plays the unifying role played by
finite-dimensional canonical Hamiltonian systems in conventional statistical
mechanics.  Thus we expect our approach to apply to Vlasov-Poisson dynamics,
kinetic theories of stellar dynamics,  drift-wave plasma models, and other
single-field models that possess the noncanonical Poisson bracket of
(\ref{ncbkt}).  Generalization to multi-field models such as reduced
magnetohydrodynamics, stratified fluids, and a variety of physics models
governed by generalization of the Poisson bracket [\cite*{thiffeault00}] of
(\ref{ncbkt}) provides an avenue for further research.

\begin{acknowledgments}
One of us (PJM) was supported by the US DOE Grant DE--FG03--96ER--54346.  Two
of us (SJ and HLS) were supported by ONR Grant N000140410282, and one of us
(SJ) also acknowledges the support of the Donald D.\ Harrington Fellows Program
of The University of Texas.

\end{acknowledgments}

\end{document}